\documentclass[%
 reprint,
 amsmath,amssymb,
 aps,
floatfix,
]{revtex4-2}

\usepackage{graphicx}
\usepackage{dcolumn}
\usepackage{bm}
\usepackage{hyperref}

\usepackage{makecell}
\usepackage{verbatim}
\hypersetup{
    colorlinks=true,
    linkcolor=blue,
    filecolor=blue,      
    urlcolor=blue,
    citecolor=blue,
    pdftitle={Overleaf Example},
    pdfpagemode=FullScreen,
    }

\def \M {\mathcal{M}} 
\def \A {\mathcal{A}} 

\newtheorem{theorem}{Theorem}
\newtheorem{definition}{Definition}

\usepackage{todonotes}
\usepackage{verbatim}
\usepackage{makecell}

\begin{document}

\preprint{APS/123-QED}

\title{Global dynamical structure from infinitesimal data}

\author{Benjamin McInroe}
  \email{Corresponding Author: bmcinroe@seas.upenn.edu}
 \affiliation{University of Pennsylvania}
 \affiliation{University of California, Berkeley}
\author{Robert J. Full}%
\affiliation{University of California, Berkeley
}

\author{Daniel E. Koditschek}
\affiliation{University of Pennsylvania}%

\author{Yuliy Baryshnikov}
\affiliation{University of Illinois, Urbana-Champaign}%

\date{\today}

\begin{abstract}
Scientists and engineers alike target modeling of complex, high dimensional, and nonlinear dynamical systems as a central goal. Machine learning breakthroughs alongside mounting computation and data advance the efficacy of learning from trajectory measurements. However scientifically interpreting data-driven models, e.g., localizing attracting sets and their basins, remains elusive. Such limitations particularly afflict identification of system-level regulatory mechanisms characteristic of living systems, e.g., stabilizing control for whole-body locomotion, where discontinuous, transient, and multiscale phenomena are common and prior models are rare. As a next step towards theory-grounded discovery of behavioral mechanisms in biology and beyond, we introduce VERT, a framework for discovering attracting sets from trajectories without recourse to any global model. Our infinitesimal-local-global (ILG) pipeline estimates the proximity of any sampled state to an attracting set, if one exists, with formal accuracy guarantees. We demonstrate our approach on phenomenological and physical oscillators with hierarchical and impulsive dynamics, finding sensitivity to both global and intermediate attractors composed in sequence and parallel. Application of VERT to human running kinematics data reveals insight into control modules that stabilize task-level dynamics, supporting a longstanding neuromechanical control hypothesis. The VERT framework promotes rigorous inference of underlying dynamical structure even for systems where learning a global dynamics model is impractical or impossible.
\end{abstract}

\maketitle

\section{Introduction}
Ever more ubiquitous computation and burgeoning data sets promise to reveal the mechanisms that underlie the behavior of complex dynamical systems directly from observation data. Data-driven modeling is particularly valuable for biological and complex engineered systems.  Their macroscopic behaviors emerge from coupled, nonlinear, and closed-loop dynamical interactions between large numbers of degrees of freedom (DoFs), themselves carrying dynamics that typically  span multiple spatiotemporal scales and levels of organization. However, the elucidation of mechanisms - in particular, the discovery of dynamic and control processes and subsystems that give rise to and stabilize task-level behavior - remains a significant challenge. Experimental limits to state observability and input excitations can present significant obstacles to dynamics learning methods whose underlying computational models may often impose inductive priors that reach beyond the scope of scientifically available hypotheses. 

To take the next step towards a data-driven paradigm for hypothesis informed discovery in general complex systems data, we introduce an \textit{infinitesimal-local-global} (ILG) framework that uses only generic (linear, infinitesimal) models of the dynamics to infer the persistent attracting state space structures that guide the global asymptotic system behavior. As a concrete application setting of broad interest, we focus our presentation on prospects for elucidating dynamics and control principles underlying agile and energetic legged locomotion behaviors, where characteristically high dimensional, discontinuous (hybrid), and nonlinear dynamics present significant obstacles to standard modeling approaches \cite{full1999templates}. However, the theory and algorithms we develop are general, assuming only necessary conditions for local structural stability of the attractor, and we thus expect broad applicability of our framework to data-driven analysis of complex systems.

Linear \cite{stephens2008dimensionality, ciocarlie2009hand} and nonlinear \cite{berman2014mapping, drnach2018identifying, deangelis2019manifold, luxem2022identifying} dimensionality reduction can reveal spatial patterns and latent variables that capture variance, but do not explicitly account for the temporal structure of trajectory datasets and are thus insufficient for assessing, e.g., topological invariants of the underlying flow and their stability. Both can fail to capture structure necessary for understanding control mechanisms \cite{yan2020unexpected}. Recent breakthroughs in applied machine learning have seen the development of sophisticated approaches that conceive trajectories as the output of a hidden dynamical system, and then fit a model that minimizes a system-specific prediction error. Approaches that parameterize the dynamics as linear modes, sets of nonlinear basis functions, or deep neural networks have demonstrated impressive prediction accuracy by incorporating insights from Koopman theory \cite{williams2015data, mezic2005spectral}, symbolic regression \cite{schmidt2009distilling, brunton2016discovering}, and sparse regularization \cite{brunton2016discovering, champion2019data}. Spectral submanifold theory can be used to learn reduced models of nonlinear dynamics as extended normal forms, demonstrating significant improvements to prediction accuracy for mechanical oscillations \cite{cenedese2022data}. 

This synthesis of machine learning and dynamical system theory has demonstrated promise for enabling discoveries in physical systems. However, data-driven insight into the transient-rich and non-stationary dynamics typical of biological systems has remained relatively elusive. Inductive biases from invariant manifold theory have enabled accurate learning of reduced models in multiscale systems with transients when a full order model is available in addition to observations \cite{otto2023learning}. Heuristics can be used to segment trajectories or the model domain into temporal windows that are well-approximated by linear systems \cite{del2003decomposition, costa2019adaptive} or spatial subdomains that are amenable to nonlinear approximations \cite{floryan2022data}. However, transitions between dynamical modes can reflect important mechanisms such as changing of constraints or internal control processes rather than limitations of the modeling apparatus\footnote{For example, the making and breaking of contact by appendages during animal and robot locomotion induces transitions between characteristically nonlinear dynamical regimes, motivating the necessity of considering infinitesimal stability properties in determining where qualitative changes in behavior occur.}. Principled tools for inferring dynamical structure from trajectory data without recourse to fitting a global model would extend the range of settings where explicit model learning could be applied, in addition to providing quantitative insights into systems where learning an explicit dynamics model is impractical.

Leveraging insights from differential topology \cite{hirsch2012differential} applied to the structure of hyperbolic invariant manifolds \cite{hirsch1977invariant}, we introduce the Vielbein Recovery of Templates (VERT) framework, an ILG paradigm for localizing persistent attracting submanifolds in output (observation) coordinates (Fig. \ref{fig:overview}). The infinitesimal component manifests itself in the assumption on the jets of the underlying dynamics, stipulating a decomposition of the tangent space into (integrable) distributions of contracting and neutral subspaces, which can be inferred from local data. The local existence of the integral submanifolds is phenomenologically strengthened to the existence of the global low-dimensional attractors and attendant basins of attraction \cite{guckenheimer2013nonlinear} \footnote{We remark that the VERT framework is applicable in hyperbolic dynamics where local contracting/expanding with respect to the offered Riemannian norm occurs only in some parts of the attractor (they need to occur {\em somewhere}, however, to satisfy the global hyperbolicity assumptions \cite{smale1967differentiable}). Once detected there, the normal hyperbolicity in broader regions can be inferred by extension along the trajectories. As an example, we cite van der Pol oscillators (whose entire limit sets satisfy the global criteria for hyperbolicity of \cite{hirsch1977invariant}, but lose a local attraction property in the standard Riemannian norm in the regions where the energy is pumped into the system by the “Coriolis term”).}. This broad framework shares with much of the prior literature the widely held premise that dynamical behavior of physical interest is typically characterized by a collapse of dimension – the presence of a low-dimensional attracting invariant submanifold onto which a system’s trajectories quickly converge within the nominally high dimensional physical state space. 

Motivated by this insight, our central contribution is the development of a theory and computational procedure for isolating from observed trajectory data the global locus of the lower dimensional attracting submanifolds that carry the long term system behavior. Critically for biological applications, our approach achieves this analysis while circumventing the need to fit a global (or piecewise global) model of the vector field to the trajectory data. Specifically, we show that a 'fiberwise distance estimator' whose near-vanishing sublevel sets indicate trajectory segments on or close to an attracting set can be constructed from infinitesimal-scale, linear dynamics estimated from local samples in a pointwise manner, affording localization of these loci of characteristic behavior without prior commitment to any parameterized model or structural assumptions beyond normal hyperbolicity --- a necessary condition for their persistence under perturbations. Using models of both continuous and hybrid dynamical systems, we demonstrate the efficacy of our approach for discovering structurally stable attracting submanifolds with a range of geometric and dynamical properties without recourse to any other prior assumptions about the nature of the full system or attractor. Finally, we illustrate the applicability of our framework to postural trajectory data from motion capture of human treadmill running whose future systematic pursuit promises to reveal insight into the neuromechanical control of the hybrid, nonlinear dynamics underlying sensorimotor behavior.

\section*{Background}
\subsection*{State Space Structures of Multiscale Dynamical Systems}

\begin{figure*}[t]
\centering
\includegraphics[width=1.0\linewidth]{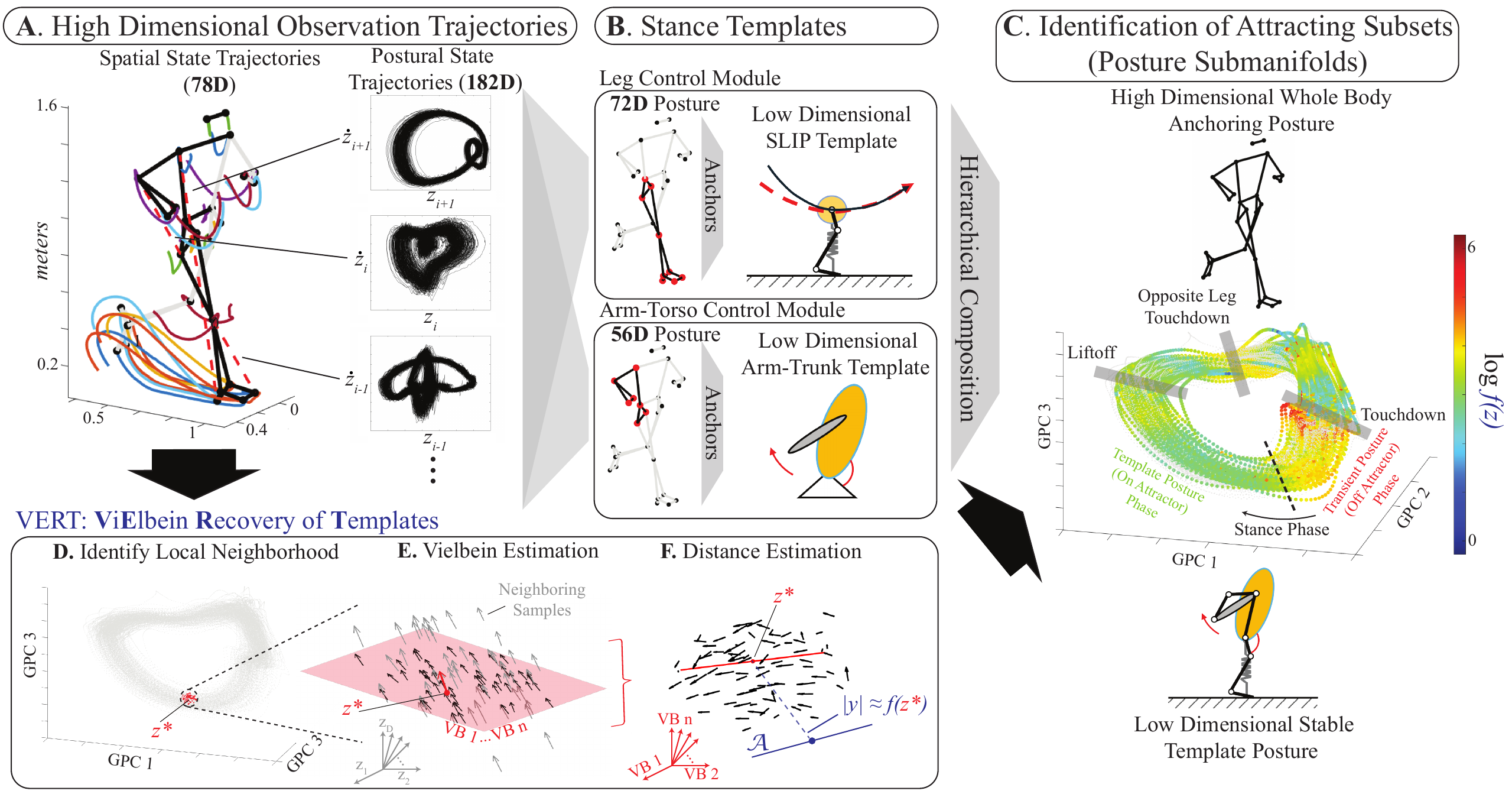}
\caption{Overview of the VERT computational framework. Panels (\textbf{A}-\textbf{B}) depict the relation to motivating hierarchical control hypotheses for complex, multi-scale systems, viewed through an illustrative application setting of the postural dynamics of human running. Panels (\textbf{D}-\textbf{F}) outline the steps of the VERT algorithm. \textbf{A.} Stable task-level system behaviors such as running emerge from the coupled, highly nonlinear, closed loop dynamics of many body segments and their interaction with the environment. \textbf{B.} We hypothesize that for many tasks, the global task-level behavior may be decomposed into hierarchically arranged control modules, whose coordinated interactions characterize the dynamics of appendages and body segments. The control module state spaces are embedded as submanifolds (postures) that carry the constituent dynamics (templates) of the full, coordinated system.  \textbf{C.} We can use VERT to learn a spatiotemporal filter on the trajectory data that isolates locally stable (template) behavior from transient behaviors reflecting disturbances and stabilizing control (anchoring). \textbf{D.} VERT estimates the distance from each sample to the attractor guiding its asymptotic behavior ($f(z)$, Def. \ref{def:sensingfunction}). The estimation procedure for a point $z^*$ begins with identification of its metric nearest neighbors in state space. \textbf{E.} A Cartesian basis for the tangent space at $z^*$ of the manifold containing the trajectories is estimated. The local subsample of the vector field (temporal differences, gray arrows) is projected onto the $n$-dimensional local vielbein basis (black arrows). \textbf{F.} Using this representation, we can estimate the component of the vector field on the vertical subspace, the local trivialization of the global, typically nonlinear vertical subbundle of the attractor. We can use this estimate to approximate the distance from $z^*$ to its base point on the unknown attractor $\A$. Steps \textbf{D}-\textbf{F} are repeated for an appropriate number of samples. Filtering the trajectory data for sublevel sets of $f(z)$ reveals the posture carrying the template dynamics (blue) and hierarchical stabilization (anchoring) intervals of what subsequent hypothesis-informed analysis can  posit as the stance leg and upper body posture during stance (See Fig. \ref{fig:human}).}
\label{fig:overview}
\end{figure*}

We consider the setting in which one has collected a dataset of trajectories from a continuous dynamical system. A common hypothesis is that the trajectories of nominally high dimensional and complex dynamical systems lie on or near a lower dimensional, often nonlinear space of latent state variables in the full order observation space, reflecting the presence of, e.g., internal constraints, symmetries, and feedback control. In the context of dynamical systems with multiple spatiotemporal scales, the low dimensional latent submanifolds are typically assumed to be invariant submanifolds of the flow \cite{kuehn2015multiple}. Using observations to discover latent invariant submanifolds and characterize their dynamics is a central goal of data-driven modeling. By capturing the global structure of the state space, learning algorithms that explicitly infer underlying invariant sets are shown to have superior predictive accuracy, even generalizing to situations where the system is subject to external forces not present in the training data \cite{haller2022dynamics}.

In principle, one could infer the invariant submanifolds from data by studying the spatiotemporal statistics of the corpus of measured samples. However, this approach implicitly assumes that the samples are in fact drawn (with noise) from the submanifold. While this assumption is justified when we know \textit{a priori} that the measurements came from steady-state behavior, it does not hold in general for many of the energetically open, nonlinear dynamical systems encountered in nature. Specifically, the presence of transient phenomena resulting from, e.g., external disturbances, noisy internal feedback control processes, and discontinuous events such as the making and breaking of physical contacts will perturb the state away from the latent submanifold and introduce transient dynamics distinct from the on-manifold behavior we want to model.

Because we are interested in experimental observations of physically relevant latent submanifolds, we assume robustness to perturbations in the form of \textit{structural stability}. The global topology of structurally stable vector fields is invariant to perturbations of the vector field \cite{smale1967differentiable}, implying that the attracting sets are robust and thus observable even in the presence of noise. Normal hyperbolicity is a sufficient condition for local structural stability; for modeling purposes, we thus assume that detectable attracting sets are \textit{normally hyperbolic invariant manifolds} (NHIMs) \cite{hirsch1977invariant}. NHIMs are important to many fields of science and engineering \cite{wiggins2013normally}, and are central to the modeling of biological locomotion and synthesis of biologically inspired robot controllers \cite{sharbafi2017bioinspired}. Further discussion of the use of hierarchically composed invariant manifolds in robotics and connections to biological motor control is included in Section 1 of the supplement.

\subsection*{Attracting Submanifolds and Dynamic Locomotion}

To realize concrete potential applications of our approach, we consider models of multiscale postural dynamics from dynamic legged locomotion \cite{dickinson2000animals}. A longstanding question in the sensorimotor control of locomotion is how the abundance of muscles, joints, and segments of the body are coordinated to achieve a large and diverse set of typically lower dimensional whole body motions. The pioneering work of Nikolai Bernstein \cite{bernstein1967coordination} suggests that whole body movements are constructed generatively from modular 'submovements'. Modern electromyography studies in both humans and non-human animals support the hypothesis that the neuromuscular system may be organized into modular units (control modules) that can be recruited in series and parallel to achieve high level motor tasks \cite{ting2015neuromechanical}. Development of a systems-level understanding of how such control modules enable robust whole-body behavior is a grand challenge of organismal biology \cite{schwenk2009grand}, with further applications to robotics and rehabilitation science.

We consider models of multiscale closed loop dynamical systems originally developed at the interface of organismal biology and legged robotics. In this domain, the pairing of a minimal low dimensional latent submanifold and its restriction dynamics with the higher dimensional dynamical system in which it is embedded is called a 'template-anchor pair' \cite{full1999templates}. A typical representation of the (high dimensional) anchor is the full postural state space of an articulated body, consisting of all joints, segments, and actuators. The corresponding (low dimensional) template consists of latent state variables whose spatiotemporal structure and behavior fully characterize the task-level dynamics, e.g., the motion of the center of mass (CoM). A canonical example is that the whole-body dynamics of running are often dynamically consistent with a simple spring-loaded inverted pendulum (SLIP) dynamical system. Here the latent state variables are the generalized coordinates for the effective length of the stance leg and its orientation relative to the ground. Experiments have shown that these latent state variables are sufficient to predict the trajectory of the CoM in animals across a range of and body sizes and morphologies \cite{blickhan1993similarity}, and can be used constructively to stabilize dynamic running behaviors in robots with 2, 4, and 6 legs \cite{holmes2006dynamics}. 

\subsection*{Hierarchical Composition of Multiscale Dynamics}

To aid in systematic analysis of hierarchical dynamical systems and their state space structures, we refine our notion of hierarchy to the (not mutually exclusive) classes of parallel and sequential dynamical hierarchies from control theory. Compositions of simple dynamical primitives such as point attractors and stable limit cycles have been used extensively as constructive models of both biological and robotic \cite{ijspeert2013dynamical, hermus2020evidence, saveriano2023dynamic, hogan2013dynamic} motor control. There is a long history of synthesizing complex whole-body behaviors from hierarchical composition of low dimensional target dynamics, originating with the control design of the first robot capable of dynamic hopping \cite{raibert1986legged}. Parallel hierarchies generalize the template-anchor pairing of the previous section by allowing the embedding and asymptotic stabilization of multiple nested submanifolds in the full order state space. Controllers based on parallel hierarchies of low dimensional dynamics inspired by animal locomotion have been used to stabilize diverse whole body behaviors \cite{de2015parallel,de2018vertical,topping2022composition,greco2023anchoring}. More recently, parallel hierarchies have served as an effective inductive bias for learning-based controllers \cite{fawcett2022toward, huang2024hilma}. Sequential hierarchies characterize near instantaneous switches in the governing dynamics as the state is propagated in time. A typical example from biology is the transition between the stance and flight phases of running. A common instance of sequential hierarchy in robotics is the back-chaining of basins of attraction around successively targeted attracting sets to achieve goal states while avoiding obstacles, either as a design choice \cite{lozano1984automatic} or as a consequence of limited control affordance \cite{Kvalheim_Koditschek_2022}.

\section*{VERT: Vielbein Recovery of Templates}
We introduce the Vielbein Recovery of Templates (VERT) framework, an ILG approach for identifying hierarchically embedded attracting sets from trajectory data. VERT takes a set of observed trajectories and returns the fiberwise proximity of each point to the normally attracting invariant manifold (NAIM) that governs its asymptotic behavior. To establish a theoretical foundation for the computational heuristics used by VERT, we model the trajectories $\{z_{i}(t), t\in [0, T_{i}]\}^{N}_{i=1}$ as samples from the (possibly noisy) flow of a dynamical system $(\M, \xi)$, where $\M$ is a piecewise smooth Riemannian manifold of dimension $n$, and $v:\M\rightarrow T\M$ is a piecewise $C^{r\geq 1}$ vector field. We assume the trajectories are embedded in a Euclidean observation space $\mathbb{R}^D$.

In the data-driven regime, we have access only to discretely sampled trajectories, and lack the coordinate representations of the tangent and normal bundles typically necessary to estimate their restriction dynamics. We thus employ insights and heuristics from the theory of normally hyperbolic invariant manifolds. Our approach is to define a {\em distance estimator} (Def. \ref{def:sensingfunction}) that evaluates proximity of a sampled state to the attracting set in the {\em fiberwise} sense. We proceed by establishing modeling assumptions necessary to analyse the behavior of the fiberwise distance estimator.

Let $\A \subset \M$ be an invariant submanifold of dimension $a$ for $(\M, v)$: $\phi_t(\A) \subset \A, t>0$, for $\phi_\cdot$ being the flow associated to the vector field $v$. We will denote by $U\supset \A$ a tubular neighborhood of this submanifold \cite{hirsch2012differential} which is attracted to $\A$. Normal Hyperbolicity implies that the trajectories starting in the neighborhood $U$ converge to $\A$, exponentially fast, uniformly in some fixed Riemannian structure. We can introduce, locally, a privileged coordinate system $z = (x,y)$ on $U$, such that the the coordinate $x$ parameterizes the points of $\A$, while the attractor is given by $\A = \{y = 0\}$. The fibers $\mathcal{F}_x$ of the projection $(x,y)\mapsto x$ are the analogues of the {\em stable manifolds} for the hyperbolic dynamics, trivialized through the selected coordinate system. We notice that this coordinate system does not have to (and in fact, almost never does) extend to a global one. However, its usage provides a convenient modeling setting, and leads to the heuristic assumptions outlined below \footnote{We use the German term \textit{Vielbein} for frame, a local coordinate system to avoid contamination with the video processing context: the term is used in this meaning in the physics literature.}.

Fix a vicinity $\A_U$ in the attractor, over which our coordinate system is defined, such that the tangent spaces to fibers at the points $(x,0)$ are exponentially contracted. Now we will explicitly state our heuristic assumptions by expressing a version of normal hyperbolicity in local coordinates $(x,y)$. The vector field $v$ describing the dynamics of the system can be represented, near a point $x_*\in \A_U$ as
\begin{equation}
\label{eq:vectorfield}
v_{x}= 
\begin{pmatrix}
    u_{0} + S(x,y)\\
    A_{1}y + T(x,y)
\end{pmatrix} = 
\begin{pmatrix}
    u_{0}\\
    A_{1}y
\end{pmatrix} + B(x,y).
\end{equation}

Here $u_{0}=v(x_*,0)\in T_{x_*}\A$ is the value of the vector field at $(x_*,0)\in U$, and $S(x,y)$ captures the vanishing at $(x_*,0)$ {\em horizontal} component of the vector field $v$. Similarly, $A_{1}y + T(x, y)$ captures the split of the {\em vertical} component into the linear (in $y$) and non-linear parts. We remark that in the standard formalism of hyperbolic sets \cite{smale1967differentiable}, the term $S$ depends on $x$ only, but we will not need this constraint. The mapping $B$ describes the residual components of the vector field after the minimal approximations of (slow) motion along the attractor and exponential convergence towards it are extracted  (nonlinearities, noise, cross-terms, etc). While we do not have direct access to these terms (as they depend on the coordinate system $(x,y)$, knowledge of which is equivalent to recovery of both the attractor and the stable foliation), this model of the infinitesimal behavior of the vector field can be used to establish a procedure to estimate the proximity of a point, $z_*=(x_*, y_*)$ to $\A$. Let $H:=D_{z}v$ and $G:=D_{z}B$ be the Jacobian matrices of mappings $v$ and $B$, respectively. 

\begin{definition}
\label{def:sensingfunction}
The {\em fiberwise distance estimator} is given by
    \begin{center}
    $f(z):= \vert H(z) v(z) \vert$
    \end{center}  
\end{definition}

We note that the Jacobian $H$ does not depend on the choice of the selected coordinate system, and that the determination of $G$ relies on the local identification of the vicinity of $z_*$ with a real linear space, but not on a specific splitting of the space into $x$ and $y$ factors. To understand the behaviour of the distance estimator $f$ notice that in the ideal, linear setting, the local vector field is given by Eq. \ref{eq:vectorfield} with $B(y)=0$. In this case, by inspection, the fiberwise distance estimator is equal to $\vert A_{1}^{2} y_* \vert$, so that $\hat{f}(z_*)$ grows linearly with the distance from $z_*$ to $\A$. Of course, the presence of noise and nonlinearities presents an obstacle to this reasoning. We proceed by showing that the behavior of $f(z_*)=f(x_*,y_*)$ with respect to $\vert y_* \vert$ can still be bounded so long as the residual component of the vector field is small relative to the hyperbolic component. We quantify this concept by introducing several constants, depending on the relative strength of the hyperbolic contraction.

\begin{definition}
    Let $C_G:=\sup\limits_{(x,y)\in U} \Vert G(x,y)\Vert$ be the upper bound on the operator norm of $G$. Further, consider the bounds on $C_+:=\sup\limits_{x\in \A_U} \Vert A_1(x)\Vert$ the operator norms of $A_1$ and $C_-:=\left(\sup\limits_{x\in \A_U} \Vert A_1(x)^{-1}\Vert\right)^{-1}$ on the reciprocal  of the operator norm of $A_1^{-1}$ (so that $C_-|y|\leq |A_1(x)y|\leq C_+|y|$ for $\in \A_U)$. Further, let $v_0:=\sup\limits_{x\in \A_U} |v(x,0)|$.
\end{definition}
Conceptually, $C_G$ describes how strong the vector field $v$ deviates from local linearity in $y$ and constancy in $x$. The constant $C_{-}$ expresses the strength of the contraction towards the attractor. Ideally, $C_G$ is small and $C_-$ is large. Note, however, that $C_-\leq C_+$.

\begin{definition}
\label{def:errorbound}
The error bound is given by
    \begin{center}
    
    \;$E(\rho) = C_G((2C_+ + C_G)\rho + v_0)$

    \end{center}
    
\end{definition}

Observe that the error bound grows quadratically with $C_G$.

\begin{theorem}
\label{thm:thm1}
In the notation of Definitions \ref{def:sensingfunction} and \ref{def:errorbound},

\begin{center}
$C_-^2\vert y \vert - E(\vert y \vert) \leq f(z) \leq C_+^2\vert y \vert + E(\vert y\vert)$
\end{center}
\end{theorem}

We provide a proof of Theorem \ref{thm:thm1} in Section 3 of the supplement. The key implication of this result is that for sufficiently small $C_G$, the estimator $f$ will be away from $0$ together with the distance $|y_*|$ to the attractor, in a quantifiable manner, allowing one to ascertain the proximity to it. We remark that the definition, and therefore the estimation and computation of $f$, requires exclusively linear operations on local data in the neighborhood of a point, and does not depend on the split of the coordinate system into $x$ and $y$, despite the fact that the formulation of our estimate does. By computing $f$ for all (or an appropriate subsample) of the trajectory data, and then filtering the $f$-sublevel sets of the full trajectory dataset (geometrically, a \textit{point cloud} in $\mathbb{R}^D$), we can isolate the set of trajectory segments that lie on or quantifiably near any physically relevant attractor $\A$ in the \textit{global} embedding (observation) space coordinates $\mathbb{R}^D$.

We detail algorithms for computing $f(z)$ on measurement data in Section 4 of the supplement. The computation is, at a high level, performed as follows. First we estimate a local Cartesian basis for $T_{z_*}\M\simeq T_{x_*}\A \oplus \mathbb{R}^{n-a}$ and compute the restriction of the local vector field to this subspace $v_{z_*}$. Local PCA, performed on the subsample of points lying within an open Euclidean ball of radius $r$ centered at $z_*$, provides a simple approach for computing a local basis of $T_{z_*}\M$ such that estimators for the tangent space and intrinsic dimension enjoy strong formal guarantees \cite{lim2021tangent}. The local sample of the vector field $V_{z_*}:\mathbb{R}^D \rightarrow T\mathbb{R}^D$ can be approximated from finite differences. The local representation of the ($n$-dimensional) $v_{z_*}$ can then be found by orthogonal projection of the ($D$-dimensional) $V_{z_*}$ onto the vielbein basis. The Jacobian $H(z_*)$ can then be estimated by centering the local vector field data and using linear estimation techniques; for experiments in the following sections, we use least squares regression with the standard MSE residual. 

\begin{figure*}[t]
\centering
\includegraphics[width=0.8\linewidth]{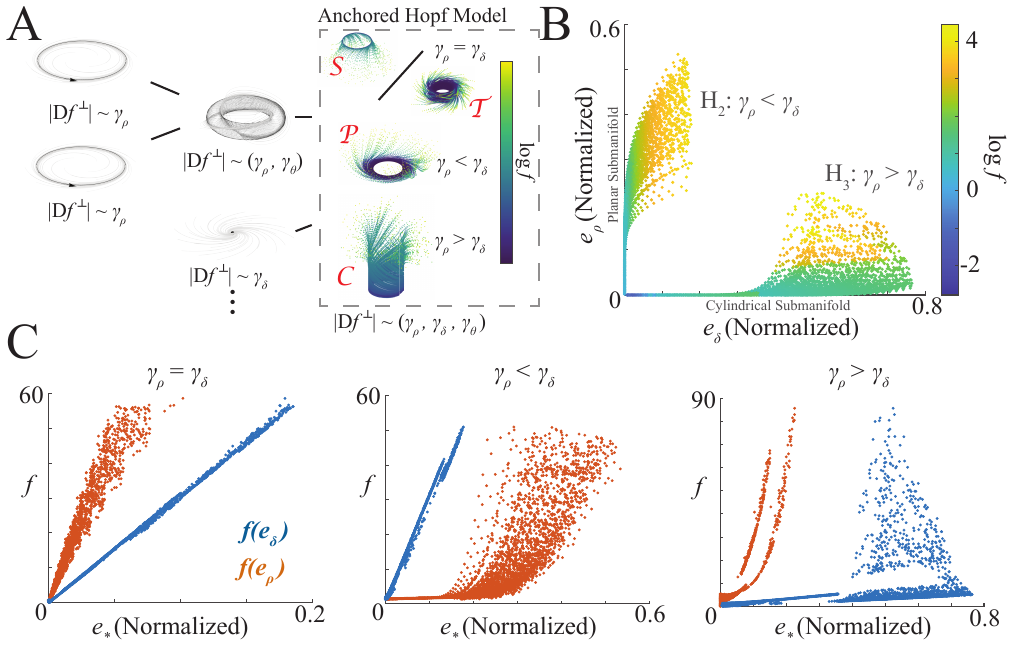}
\caption{Application of VERT to the A-HM model illustrates the mechanisms laid out in Fig. \ref{fig:hopfresults} that underlie the hypotheses in Table \ref{table:hypotheses}: the distance estimator is sensitive to the locally dominant dynamics at the sampled point. 
\textbf{A.} Hierarchical composition of the A-HM (Eqns. \ref{eq:hopf} and \ref{eq:anchor}). The cross product system of two limit cycles produces a limit torus, whose product with a higher dimensional stable spiral generates qualitatively distinct mechanisms for anchoring the global attractor. \textbf{B.} Visualization of VERT estimator ($f$) in the error coordinate plane. Observe the initial alignment of the $f$-level sets with the error coordinate of the stronger subsystem. Upon reaching an intermediate attractor (at vertical and horizontal axes), $f$ continues to decay more slowly to its global minimum. \textbf{C.} Sensitivity plots of $f$ for different relative subsystem gains. When $\gamma_\rho = \gamma_\delta$, $f$ varies linearly with both error coordinates. When one gain is larger, $f$ shows heightened sensitivity to the stronger system.}  
\label{fig:hopfsensitivity}
\end{figure*}

\begin{figure*}[t]
\centering
\includegraphics[width=1.0\linewidth]{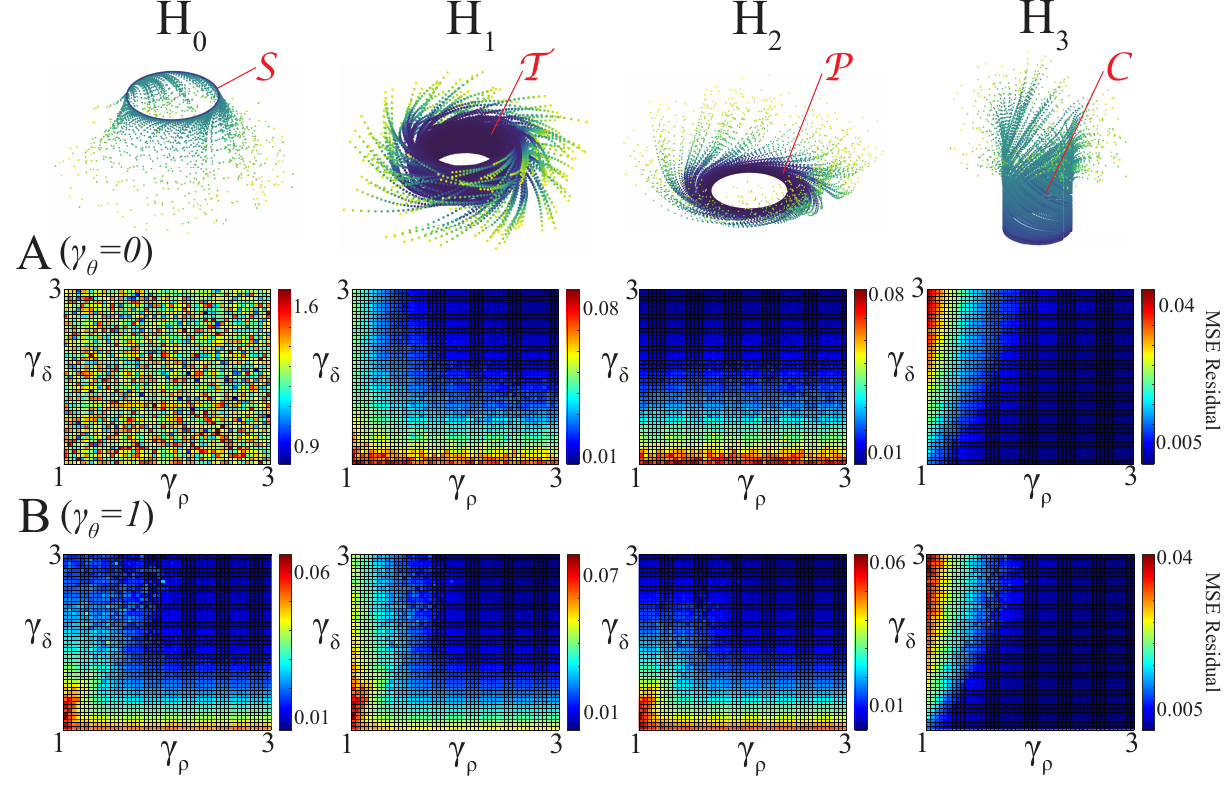}
\caption{Parameter sweep experiments illustrate sensitivity of the VERT estimator to attractor hierarchies. See Table 1 for hypothesis definitions \textbf{A.} When $\gamma_\theta=0$, the limit cycle $\mathcal{S}$ does not exist, and the outputs of VERT accordingly support rejection of $H_0$. $H_{1-3}$ are supported, as the residual decreases with both $(\gamma_\rho, \gamma_\delta)$ for $\mathcal{T}$, $\gamma_\delta$ for $\mathcal{P}$, and $\gamma_\rho$ for $\mathcal{C}$. \textbf{B.} When $\gamma_\theta=1$, $\mathcal{S}$ exists and is detected. $H_{0-3}$ are supported, with similar behavior to the $\gamma_\theta$ for the intermediate submanifolds.}
\label{fig:hopfresults}
\end{figure*}

\section*{Results}

\subsection*{VERT is Sensitive to Attractor Hierarchies}
\label{sec:hopfresults}

Detecting the presence of subprocesses evolving on distinct time scales is critical for model inference and interpretation of systems constructed by parallel composition of stable subsystems. We hypothesize that the ILG nature of the VERT pipeline affords sensitivity to both global and intermediate attracting sets through a detectable change in the behavior of the distance estimator along a trajectory as it transitions between them. We construct a model system inspired by dynamic movement primitives, the anchored Hopf model (A-HM). The A-HM embeds a pair of coupled Hopf-type limit cycles in a higher dimensional Euclidean space via a hyperbolic linear flow. The A-HM template-anchor system is easiest to render in polar coordinates. The coupled limit cycle subsystem is:

\begin{equation}
\label{eq:hopf}
    \begin{split}
    &\dot{\rho}_i = \gamma_\rho (1-\rho_i)\rho_i \\
    &\dot{\theta}_i = \omega + \gamma_\theta \sin(\theta_i - \theta_j)
    \end{split}
\end{equation}

where $i,j\in \{1, 2\}$. We use the variable $x \in \mathbb{R}^{D}$ to refer to the corresponding Euclidean coordinate. Observe that the dynamics of the phase variables decouples from the amplitudes $\rho$, and has the structure of a Kuramoto oscillator system. The system entrains to a global limit cycle so long as $\gamma_\theta>0$. The auxiliary embedding subsystem has the form:

\begin{equation}
\label{eq:anchor}\dot{y}=Ay
\end{equation} 
with a Hurwitz $A$, and $y=(y_5,\ldots,y_D)$ is the vector of the auxiliary variables. The state vector for the full system composed of Eqs. \ref{eq:hopf} \& \ref{eq:anchor} is thus $(z_1,...,z_D)=(x_1,...,x_4,y_5, ..., y_D)\in \mathbb{R}^{D}$. The rate of convergence to the anchored plane at $y=0$ is set by the real part of the eigenvalues $\sigma$ of $A$, $\Re\sigma = -\gamma_\delta$. 

We test the sensitivity hypotheses elaborated in Table 1 using simulated trajectories from the A-HM, while systematically varying the control parameters $\gamma_\rho$ and $\gamma_\delta$, which determine the rate of convergence for the two subsystems. We define the error coordinates $e_\rho$ and $e_\delta$ as the normalized Euclidean distance from a point on the trajectory to the limit torus of Eq. \ref{eq:hopf} and limit point of Eq. \ref{eq:anchor}, respectively. These error coordinates provide ground truths for the global fiberwise distances that VERT estimates. The qualitative behavior of $f$ across the $(\gamma_\rho, \gamma_\delta$) plane is illustrated in Fig. \ref{fig:hopfsensitivity}. In these experiments we set $\gamma_\theta=0$ without loss of generality, so that the limit torus is the ultimate global attracting submanifold. Trajectories can either converge directly to the limit torus $\mathbb{T}^2$, or first approach either the plane of the limit cycle subsystem ($\mathcal{P} \cong \mathbb{R}^4$) when $\gamma_\delta>>\gamma_\rho$ or the space $\mathcal{C} \cong \mathbb{T}^2 \times \mathbb{R}^{16}$ when $\gamma_\rho >> \gamma_\delta$, before converging to the limit torus. We quantify proximity with the normalized error coordinate $e_{\star}$, and observe empirically that when $\gamma_\delta > \gamma_\rho$, $\frac{\partial f}{\partial e_\delta} > \frac{\partial f}{\partial e_\rho}$ in the near-field of the attractor and vice-versa, indicating the switch in sensitivity of the estimator when the relative magnitudes of the gains are permuted. A table summarizing these detection hypotheses and a visual rendering of the relationships between trajectories and their intermediate attracting submanifolds is included in Section 5B of the Supplement.

\begin{table}
\begin{center}
\begin{tabular}{ |c|c|c| } 
\hline
Subsystem Gain & \makecell{Anchoring Hierarchy} & \makecell{VERT Detection \\Hypothesis} \\
\hline
$(\mathbf{H_0})$ $\gamma_\theta \gg 0$ & $z_i \in \mathbb{R}^d \rightarrow \cdot\cdot\cdot \rightarrow\mathcal{S} \ni z_f$ & Detect $\mathcal{S}$ \\ 
\hline
$(\mathbf{H_1})$ $\gamma_\rho \approx \gamma_\delta$ & $z_i \in \mathbb{R}^d \rightarrow \mathcal{T} \rightarrow \mathcal{S} \ni z_f$ & Detect $\mathcal{T}$ \\ 
\hline
$(\mathbf{H_2})$ $\gamma_\rho \ll \gamma_\delta$ & $z_i \in \mathbb{R}^d \rightarrow \mathcal{P} \rightarrow \mathcal{T} \rightarrow \mathcal{S} \ni z_f$ & Detect $\mathcal{P}$, $\neg \mathcal{C}$ \\ 
\hline
$(\mathbf{H_3})$ $\gamma_\rho \gg \gamma_\delta$ & $z_i \in \mathbb{R}^d \rightarrow \mathcal{C} \rightarrow \mathcal{T} \rightarrow \mathcal{S} \ni z_f$ & Detect $\mathcal{C}$, $\neg \mathcal{P} $ \\ 
\hline
\end{tabular}
\end{center}
\caption{Table of hypothesized anchoring mechanisms ($\mathbf{H_{0-3}}$) for the A-HM system (Eq. \ref{eq:hopf}, \ref{eq:anchor}). The choices of subsystems gains modulate which intermediate attracting submanifolds a trajectory visits while approaching the global invariant submanifold. We study the sensitivity of VERT to such hierarchical systems by testing the detection hypotheses in the right column.}
\label{table:hypotheses}
\end{table}

To develop a systematic test of the detection hypotheses, we use the continuous output $f$ to construct a binary classifier that predicts the "template" set: the trajectory segments that lie sufficiently near the globally invariant attracting set. In the context of our theory, the template set can be viewed as a finite approximation to the zero set $f^{-1}(0)$. We then quantify the quality of the template set estimate as the MSE (mean squared error) projection residual, the mean over orthogonal Euclidean projection distances from each point in the template set to the true submanifold. Lower values of the residual indicate that the predicted template set more closely conforms to the hypothesized submanifold, while relatively higher values indicate that the identified template set contains a larger proportion of samples distant from the target attracting submanifold. The results are summarized in Fig. \ref{fig:hopfresults} for experiments where the lowest level attractor is either the limit torus ($\gamma_\theta=0$) or the limit cycle ($\gamma_\theta=1$). When one of the anchoring gains $(\gamma_\rho, \gamma_\delta)$ is larger than the other, relatively lower magnitudes of the residual are achieved for the corresponding submanifold, with relatively higher residual for the submanifold corresponding to the weaker gain. We note that there are qualitative differences in the results over the $(\gamma_\delta,\gamma_\rho)$ parameter space between the experiments with and without a global limit cycle. We note that varying $\gamma_\theta$ effectively introduces a third time scale, which in turn can influence the behavior of $f$. However, the behavior of the VERT predictions over the $(\gamma_\rho, \gamma_\delta)$ parameter space is empirically consistent for both settings of $\gamma_\theta$.

\subsection*{VERT Finds Continuous Attractors in Subdomains of Hybrid Systems}
\label{sec:vdp_results}

\begin{figure*}[ht]
\centering
\includegraphics[width=1.0\linewidth]{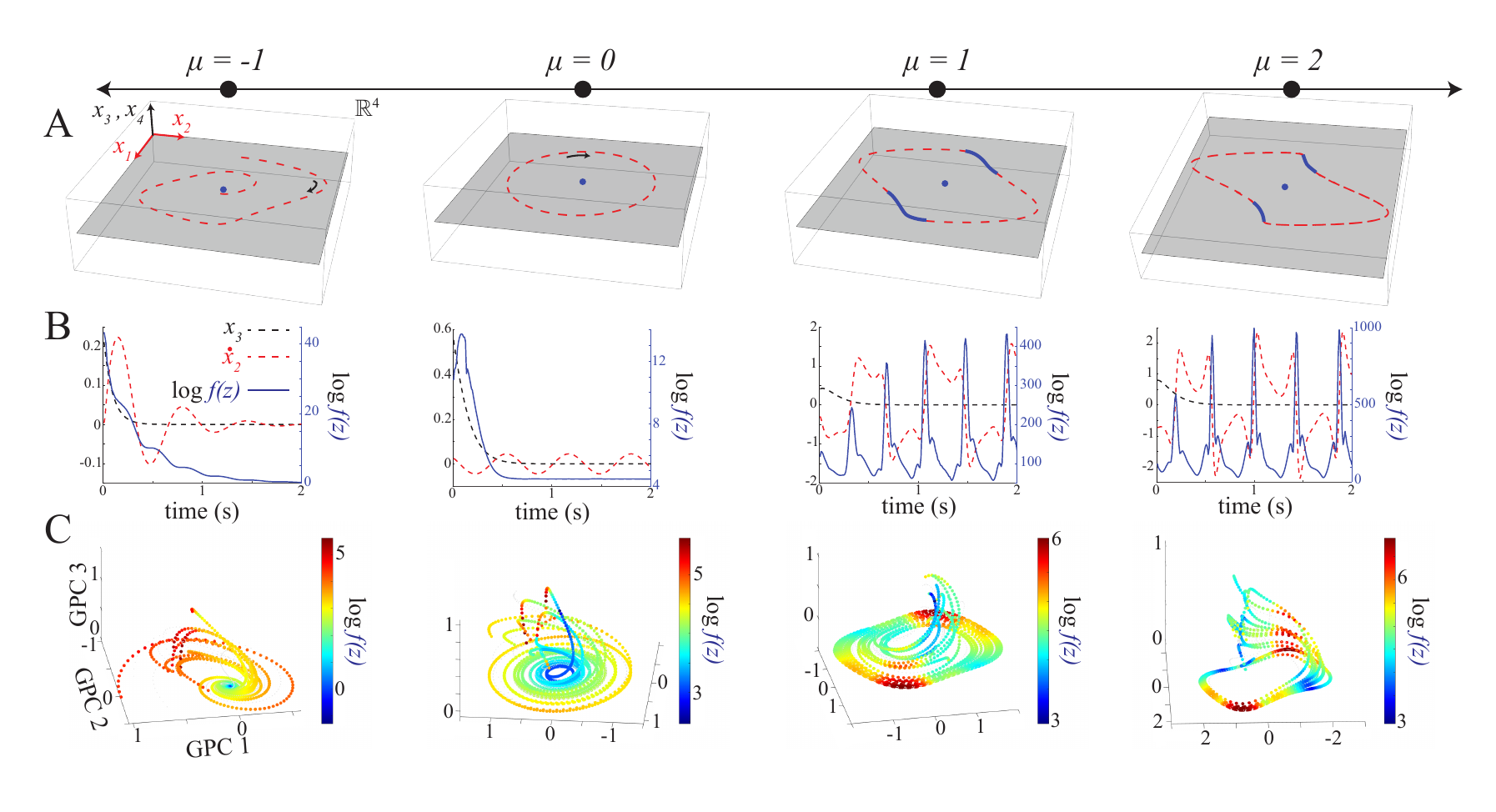}
\caption{VERT predicts locally stable regions across continuous and hybrid behaviors in the van der Pol system. \textbf{A.} Hierarchical and sequential attractors of Eq. \ref{eq:vdp} across the nonlinear damping parameter $\mu$. The grey plane $(x_1, x_2)$ contains the van der Pol vector field, and is globally attracting. Within the van der Pol plane, red curve segments depict highly impulsive regions along the attracting limit cycle while blue curve segments depict the more slowly evolving regions. \textbf{B.} Representative trajectory time series for the distance estimator $f(z)$ (blue), the position variable of the anchoring system $(\dot{x}_3, \dot{x}_4)$ (black dashed), and the acceleration of the van der Pol system $\dot{x}_2$ (red dashed). \textbf{C.} Visualizations of the trajectory datasets plotted with respect to their leading three global principal components (GPCs), and colored by magnitude of the distance estimator. Low values of the distance estimator indicate predict the locus of a locally hyperbolic attracting set. The highly impulsive episodes of the van der Pol oscillator disturb the predictions, which, otherwise, accurately locate the limit cycle locus, suggesting the distance estimator's utility for detecting hybrid behavior.}
\label{fig:vdp_results}
\end{figure*}

Detection of discontinuous or jump phenomena is necessary for inference of systems with non-smooth dynamics, such as legged locomotion or manipulation. We analyze the behavior of VERT on simulated data from a continuous system with an emergent jump behavior, termed the anchored van der Pol system (A-vdP). A-vdP is a van der Pol type oscillator anchored in $\mathbb{R}^4$ by an overdamped harmonic oscillator. In Cartesian coordinates:

\begin{equation}
    \label{eq:vdp}
    \begin{split}
    &\dot{x}_1 = x_2 \\
    &\dot{x}_2 = -\sin(x_1) + \mu(\cos^{2}(x_1) - 0.5)x_2 \\
    &\dot{x}_3 = x_4 \\
    &\dot{x}_4 = -x_3 - b x_4
    \end{split}
\end{equation}

Where $b$ is the damping coefficient for the harmonic oscillator. Depending on the choice of the continuous parameter $\mu$, the global dynamics of the A-vdP system are characterized by an attracting fixed point, Hamiltonian periodic orbits, or a limit cycle with relaxation oscillations. When $\mu$ is sufficiently large, the restriction dynamics to the limit cycle are interpreted as an effectively hybrid dynamical system with fast jumps between relatively slow hyperbolic segments \cite{van2007introduction}. In principle, the global limit cycle could be discovered from data as the locus of fixed points of the Poincare section, but this analysis would not provide insight into the jump behavior or locally hyperbolic nature of the flow on the limit cycle itself. We hypothesize that when $\mu$ is large enough that the limit cycle exhibits jump phenomena, the ILG nature of VERT will allow identification of the locally hyperbolic subdomains. 

To test this hypothesis and explore the efficacy of VERT across dynamical regimes of the A-vdP system, we use VERT to predict attracting sets from simulated data for four values of the bifurcation parameter $\mu$ (Fig. \ref{fig:vdp_results}). We compare representative predictions from VERT to the ground truth trajectories from the A-vdP system (Fig. \ref{fig:vdp_results}B). The position coordinate of the anchoring subsystem, $x_3$, goes to zero as the trajectory approaches the van der Pol subsystem plane. The acceleration of the van der Pol subsystem, $\dot{x}_2$, only approaches zero asymptotically when the attracting set is a globally attracting fixed point ($\mu=-1$). Jump behaviors on the limit cycle are accompanied by a sudden change in the sign of $\dot{x}_2$. 

We visualize the spatial behavior of VERT predictions by projecting the trajectory data onto its first three GPCs (Fig. \ref{fig:vdp_results}C). When $\mu=-1$, the locus of the attracting fixed point at the origin is apparent as the global minimum of $\log f(z)$. When the vector field in the van der Pol plane is Hamiltonian ($\mu=0$), the distance estimator takes its minimum on the van der Pol plane, which is an attracting set for the anchoring subsystem. The periodic orbits have constant energy that increases with their radius; the behavior of the distance estimator is accordingly approximately constant over individual orbits, with this steady state value decreasing with orbit energy. When $\mu$ is positive and sufficiently large, the vector field in the van der Pol plane is much stronger than the anchoring vector field. The value of the distance estimator peaks sharply at the locus of jumps on the limit cycle, and is minimal when the trajectory is near the pair of locally attracting limit points on the slow phase of the limit cycle. While the limit cycle is globally attracting, the VERT predictions reveal the locally stable and unstable segments. It is worth noting that while superficially, VERT recovers only fragments of the (exponentially) stable limit cycle, this is enough to reconstruct all of it, given the global convergence properties.

\subsection*{VERT Reveals Candidate Control Modules in Human Running Data}
\label{sec:humanresults}

\begin{figure*}[ht]
\centering
\includegraphics[width=1.0\linewidth]{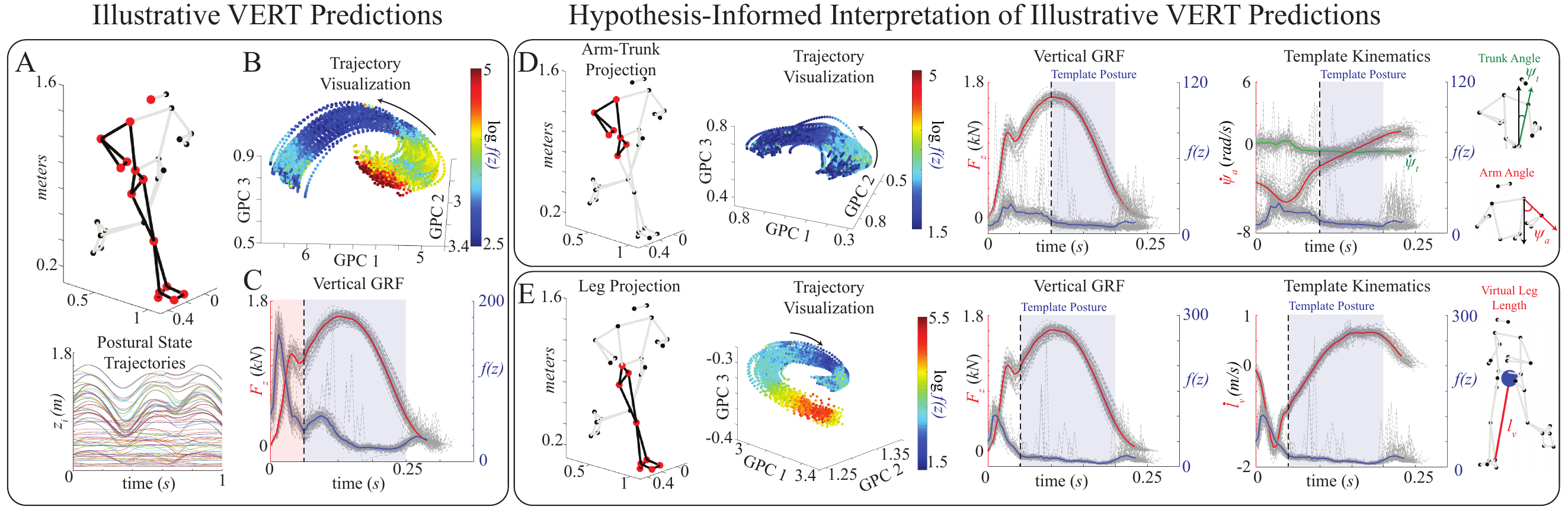}
\caption{Illustrative VERT predictions of running control modules are supported by comparison to empirically validated template models based on data withheld from the VERT pipeline. Blue traces show the VERT distance estimate. \textbf{A.} The postural state data input to VERT are presented as trajectories through the space of spatial pairwise distances (and their velocities) between body markers, $z\in \mathbb{R}^{182}$. \textbf{B.} Visualization of the stance phase postural trajectories colored by value of the distance estimator, projected onto the first 3 global principal components (GPCs) of the postural state space. Arrow indicates direction of flow, from impact to lift-off. \textbf{C.} Stance phase averaged time series (with shadowed data traces) of the distance estimator $f(z)$ (blue trace), aligned with vertical ground reaction force (GRF), $F_z$ (red trace, withheld from VERT). \textbf{D-E.} We use VERT to search for control modules in the Arm-Trunk (\textbf{D}) and Leg (\textbf{E}) postural state based only upon projections of the purely kinematic data (visualized in the lower plot of \textbf{A}). As a validation experiment probing VERT's ability to find postures carrying previously identified template dynamics (\textbf{E}), and to suggest novel postures that might carry new templates (\textbf{D}), we align and compare the averaged distance estimator $f(z)$ to the average vertical GRF time series ($F_z$) and kinematics informed by the previously known template hypotheses - the trunk and arm pitch angles ($\psi_t$ and $\psi_a$, respectively, in row \textbf{D}) and the virtual leg length from the COM to heel ($l_v$ in row \textbf{D}). The template phase of stance predicted by VERT's proposal of a stabilized posture submanifold (the blue shaded interval that marks the kinematic data at which $f(z)$ vanishes) aligns with the canonical dynamic behavior of the SLIP model \cite{full1999templates, holmes2006dynamics}.}
\label{fig:human}
\end{figure*}

The presence of low dimensional spring loaded inverted pendulum (SLIP) dynamics embedded in the high dimensional postures of running, trotting, and hopping legged animals spanning mass and length scales is a foundational principle of legged locomotion \cite{blickhan1993similarity}. Observation and identification of SLIP dynamics in data has required model-based analysis of dynamics variables including ground reaction force (GRF) and global center of mass position. The ability to detect the presence of such template dynamics from kinematics data alone would enable the discovery of new classes of dynamical locomotion templates without imposing the biases inherent in fitting a pre-determined model. However, such a procedure has remained elusive. Dynamical systems theory suggests that stability of the low dimensional dynamics requires stability of a characteristic low dimensional posture \cite{holmes2006dynamics}.

We use publicly available motion capture data of human treadmill running \cite{maus2014data} (published in \cite{maus2015constructing}) as a validation experiment to test VERT's capacity to detect template dynamics from postural kinematics measurements alone. The canonical latent variables of the SLIP embedding are the state variables of an extensible "virtual leg" extending from the center of mass to the contact foot \cite{holmes2006dynamics, poulakakis2009spring}, and the characteristic stance dynamics of an inverted spring mass system are observable from privileged measurements - GRF and COM position - available in the dataset, but withheld from VERT. 

Predictions from the VERT framework benefit from trajectory datasets containing large variation in initial states and rich time-dependent disturbances, as the primary object of estimation is the pointwise restriction of the flow to the fibers of the invariant set, rather than the flow on the invariant set itself. Treadmill running is a relatively steady state behavior, with perturbing forces arising primarily during the impulsive dynamics of foot-ground contact. However, even relatively steady state legged locomotion behaviors contain considerable variability due in part to the presence of sensorimotor noise and latency. In quantitative analyses of behavior, it is standard to reduce postural kinematic variability by projection to latent subspaces (e.g., global principle components) to isolate essential stereotyped features that are useful for downstream tasks such as segmentation and labeling \cite{brown2018ethology}. Our analysis with VERT (Fig. \ref{fig:human}) demonstrates a complementary perspective: the transient "off-manifold" dynamics of the trajectory ensemble contain valuable information about the mechanisms that stabilize the observed pattern of behavior.

Because the SLIP template model represents a single virtual leg restricted to the sagittal plane, we focus our analysis on sensors from the right side of the body during the stance phase to facilitate comparison (Fig. \ref{fig:human}A). The VERT predictions for the whole body template are shown in Fig. \ref{fig:human}B-C. Visualizing the distance estimator over the spatiotemporal projection of the trajectories onto their leading GPCs (Fig. \ref{fig:human}B) reveals a phase of transient postural dynamics immediately following impact, with closest proximity to the template posture occurring during the latter half of stance. Temporal alignment of the distance estimator with the vertical ground reaction force ($F_z$), withheld from VERT, shows that the initial post-impact transient predicted by VERT aligns with the impact transient phase of $F_z$ (Fig. \ref{fig:human}C, red shaded region), and decays near the transition to the post-transient (SLIP-like) GRF profile. Interestingly, VERT predicts the persistence of a weaker transient following the transition to SLIP-like GRF behavior. This transient decays near the functional transition from energy absorption to propulsion, where the slope of $F_z$ changes sign. Low values of the distance estimator thus align with SLIP template behavior inferred from the privileged observables (vertical GRF and virtual leg length).

These predictions support the longstanding practice of modeling running dynamics as a hybrid limit cycle system. The results of our experiments in the preceding sections suggest that we can take a step further towards identifying the responses of specific appendages and body segments (i.e., control modules) to impact by first projecting the trajectories to the corresponding subspace of the full observation space, and then applying VERT to infer convergence of the projected dynamics to the attracting posture submanifold presumed tocarry some template dynamics. We explore two control module hypotheses. The first is informed by existing template-anchor models of running dynamics: a control module for the leg posture expected to carry the SLIP template (Fig. \ref{fig:human}E). The second posits a control module for the arm and torso posture that the template/anchor framework would presum to carry some template which is, nevertheless, presently unknown (Fig. \ref{fig:human}D).

Empirical measurements of ground reaction forces and COM kinematics withheld from VERT show SLIP-like behavior only after an initial transient phase following impact. VERT's distance estimator for the leg projection converges after the $F_z$ impact transient (Fig. \ref{fig:human}E), with the second transient apparent in the full body projection absent, supporting the hypothesis that the initial postural transient results from stabilization of the SLIP posture. We estimate the length of the 'virtual leg', $l_v$, as the distance from the COM to the right medial ankle marker. We again emphasize that only postural variables were used by VERT, so information about the COM position and $l_v$ was not directly available to VERT when computing the distance estimator. Convergence of the distance estimator to its near-zero values\footnote{$f$ takes its values in the co-domain of real numbers, whereas its level sets are comprised of points in its domain of data samples.} identifies a sublevel set of data samples that coincides precisely with the transition to SLIP-like kinematics for $l_v$, supporting the validation hypothesis that VERT has accurately predicted the transition to the posture carrying the template.

Whole-body bipedal template running models that include arm and trunk dynamics are relatively less studied, in part due to the difficulty of inferring the control target of the more complex multi-body mechanical system. However, we can observe, again by comparison with the additional GRF data withheld from VERT, that its distance estimator's near-zero sublevel set of arm and trunk projected data samples determined by inspection (Fig. \ref{fig:human}D), coincides with the propulsive phase of stance (the phase following the peak of the vertical GRF). Interestingly, the template phase predicted by VERT coincides with the transition to nominally steady state kinematics of the arm posture, quantified by the spatial angle between the proximal segment of the stance arm and the gravity vector ($\psi_a$). The trunk-gravity pitch angle ($\psi_t$) also exhibits steady state behavior during the predicted template phase, although the onset occurs prior to the convergence of the distance estimator to the sublevel set, suggesting sensitivity primarily to the arm swing dynamics for the arm-trunk projection. Note that both $\psi_a$ and $\psi_t$ contain privileged information (orientation with respect to gravity) not used by VERT to compute the distance estimator. These results suggest that arm swing dynamics may have an important role for task-level control during the propulsive phase of stance, highlighting a promising direction for future study.

Finally, we note that these results complement recent experimental and simulation studies on human running showing that intrinsic variability contains critical information for explaining task-level stability mechanisms \cite{seethapathi2019step}. Even in less energetic behaviors such as manual grasping, trajectory components excluded by both linear and nonlinear dimensionality reduction contain task-specific structure that could be valuable for understanding sensorimotor control \cite{yan2020unexpected}. 

\section*{Discussion}

We have proposed VERT, a framework for using observed trajectory data to estimate the attracting sets (termed "postures") that govern their asymptotic behavior (the restriction dynamics termed a "template" Fig. \ref{fig:overview}). We show that the relative fiberwise proximity of a measured state to the locally strongest attracting set can be estimated using a model agnostic, infinitesimally (i.e., pointwise) motivated linear calculation, applied to only a spatially local subsample of the full dataset, enabling inference of the global qualitative behavior of the system without recourse to a particular global model of the dynamics. The VERT framework complements existing approaches for learning predictive dynamics models from data, which typically assume post-transient training data \cite{otto2023learning}, by providing a procedure to isolate transient trajectory segments from the full dataset. Because VERT does not require fitting a global model, this can be done in pre-processing. VERT further affords the ability to discover dynamical structure (the locus of attracting sets) from data that may be less amenable to global autoregressive model fitting due to the complexity of the underlying dynamics (e.g., presence of multiscale and discontinuous phenomena) and obstacles to model interpretation (e.g., systems that lack ground truth constitutive models). Stable attracting sets provide critical insight into the mechanisms underlying the dynamics of biological systems including locomotion \cite{holmes2006dynamics}, gene regulatory networks \cite{rand2021geometry}, and neural population dynamics \cite{vyas2020computation}. We thus expect broad utility of VERT for learning models of biological and engineering systems, where the behavior of observed trajectories is strongly influenced by persistent disturbances and internal closed loop control processes. 

We demonstrated the efficacy of VERT for accurately identifying attracting sets in synthetic data from dynamical systems with both parallel (Fig. \ref{fig:hopfresults}) and sequential (Fig. \ref{fig:vdp_results}) hierarchies, validated its ability to find posture submanifolds carrying previously discovered template dynamics from experimentally measured human running kinematics (Fig. \ref{fig:human}). Our results show that when subsystems with different characteristic time scales are simultaneously active, VERT estimates the distance to the attracting set of the locally dominant subsystem, affording sensitivity to and potentially inference of such parallel hierarchies. In systems with hybrid or jump-like behaviors, VERT detects the locally stable attracting sets on continuous subdomains of the global attracting set. In the human running data, which exhibits features of both parallel and sequential dynamical hierarchy, this sensitivity suggests inference of potential control modules during the stance phase of running. The sublevel sets of the distance estimator accurately predict the phases of stance where ground truth kinematic observables exhibit template behavior, e.g., SLIP-like leg length dynamics and vertical GRF profile. Mounting experimental evidence suggests sensitivity of muscle spindle primary afferents to the rate of change of loading force as a mechanism for reflexive stabilization of postural perturbations \cite{lin2019yank}. The VERT predictions of post-transient template phases of stance are characterized by linear or constant accelerations (i.e., zero yank) of the 'template' kinematic variables for the respective arm-torso and leg control modules. These observations support the possibility of using VERT to accurately predict reflexive muscle activation dynamics from motion capture data, a prospect with potential breakthrough applications for biomechanics, robotics, and simulation of human movement. The identification of distinct transient timescales further suggests that VERT can be used in experimental design to develop perturbation assays that untangle the roles of distributed postural control modules for stabilizing locomotion. The finite time Lyapunov exponent has served as the standard metric of dynamic stability in both human \cite{england2007influence} and animal \cite{ahamed2021capturing} locomotion. The fiberwise distance estimate afforded by VERT is fundamentally distinct from the information contained in the Lyapunov spectrum (see SI Sec. 2 for further discussion). We expect that VERT will allow researchers to take the next step towards probing the specific mechanisms that give rise to the global stability of observed locomotion behaviors.

Integration of the VERT pipeline with tools for geometric and topological inference is a natural next step --- discovered posture submanifolds likely (e.g., almost always in robotics) have dimensions greater than $1$, and are almost always highly nonlinear. This motivates developing a funnel to the computational pipes of Topological Data Analysis \cite{gtda2018}, allowing one to recognize, in a model-agnostic way, the topological structure of the underlying attracting set. As the attracting sets are still expected to be of relatively low dimensions, we conjecture that the available topological inferences (primarily, ranks of the homology groups) would be sufficient to achieve this goal. Once the underlying attracting set is recovered, the estimation of the template behavior --- the dynamics on the attractor --- could be achieved by the standard approximation tools \cite{wang2016data, yu2024learning}. The promising results from the A-HM and A-vdP models motivate further work on inferences for systems with multiscale behavior. We showed that applying VERT to hypothesis-informed projections of the human postural state can provide insight into the dynamics of particular appendages and body segments. Learning such projections directly from the outputs of VERT is an exciting prospect. To ensure our analysis of the human running data was refined to the stance phase, we used the independently measured vertical GRF signal to segment the full trial into phases when the right leg was in contact. In principle, integrating detection of hybrid reset events into VERT should be possible. The saltation matrix characterizes the first order variational update at a reset \cite{kong2024saltation}, and can be estimated from the local sample covariance matrices, which are already computed and stored by VERT to estimate the vielbein. Future work will explore the prospect of using such estimates to localize guard surfaces in state space.

\section*{Materials and Methods}
\subsection*{Hopf Dataset}
Trajectories for the A-HM model were produced by integrating the model defined by Eqns. \ref{eq:hopf} \& \ref{eq:anchor} using MATLAB's ode45 function for the integration interval $[0:0.05:10]$. 10 polar coordinate pairs were sampled from the continuous uniform distributions $\rho \sim \mathcal{U}(2,4)$ and $\theta \sim \mathcal{U}(0, 2\pi)$. For each polar coordinate pair $(\rho, \theta)$, five initial conditions were computed by sampling: $(\rho, \theta) + \mathcal{N}(0, 0.3)$, producing a dataset of 50 trajectories. Trajectories were transformed to Cartesian coordinates before passing to VERT. To produce the plots in Fig. \ref{fig:hopfresults}, control parameter combinations $(\gamma_\rho, \gamma_\delta)$ were produced from a uniformly spaced $50\times 50$ grid with $\gamma_\rho,\gamma_\delta \in [1,3]$. VERT was used to compute the distance estimator for 10 trajectories. The MSE residual was computed over these 10 trajectories. For each control parameter combination, we performed 5 trials each with independently sampled initial conditions, and the MSE residuals for each trial were averaged to produce Figs. \ref{fig:hopfresults}C,D. Details for computing the MSE residuals are described in Section 5D of the supplement. Parameters used to create the plots in Fig. \ref{fig:hopfresults} are available in Section 5C of the supplement.

\subsection*{van der Pol Dataset}
Trajectories for the A-VDP model were produced by integrating the model defined by Eqn. \ref{eq:vdp} using MATLAB's ode45 function for the integration interval $[0:0.01:20]$. To produce the plots in Fig. \ref{fig:vdp_results}, 20 initial conditions were independently sampled from the normal distributions: $x_1 \sim \mathcal{N}(0.5, 0.2)$, $x_2 \sim \mathcal{N}(0.3, 0.2)$, $x_3 \sim \mathcal{N}(0.5, 0.2)$, $x_4 \sim \mathcal{N}(0.1, 0.2)$. The numerical derivative $\dot{x}_2$ plotted in Fig. \ref{fig:vdp_results} was obtained using first order finite differences.

\subsection*{Human Running Data}
The human running kinematics dataset consisted of a single continuous 4 min. trial (S3R1) obtained from the Dryad repository \cite{maus2014data}. To parameterize the postural state of the right side of the body, we used the Euclidean distances between all pairwise combinations of the markers on the right side of the body as the position coordinates (see Section 7 of the supplement for marker labels), and the numerical derivatives of the position trajectories (obtained using finite differences) as the velocity coordinates, resulting in a state dimension of $2 \times \binom{14}{2}=182$. The postural state trajectories were passed directly to the VERT pipeline. To produce the plots in Fig. \ref{fig:human}, the vertical ground reaction force data was used to estimate right  foot touchdown and liftoff events. The data was then aligned such that touchdown events occurred at $t=0 s$. Mean trajectories for all variables over the stance phases were computed by linearly resampling the data on a uniform interval and computing the means across trials at each time step. The projected datasets were obtained following the above procedure for the marker sets used in the upper and lower body projections. 

\begin{acknowledgments}
This work was funded in part by AFOSR  HyDDRA MURI Award FA9550-23-1-0337 and in part by the University of Pennsylvania
\end{acknowledgments}
\bibliography{apssamp}


\end{document}